# COMMENTS ON 'VISCOUS DAMPING OF NON-ADIABATIC MHD WAVES IN AN UNBOUNDED SOLAR CORONAL PLASMA' BY KUMAR AND KUMAR

V.S. PANDEY and B. N. DWIVEDI

*Department of Applied Physics, Institute of Technology*
*Banaras Hindu University, Varanasi-221 005, India*
E-mail: pandey_vs@yahoo.com, bholadwivedi@yahoo.com

**Abstract.** We consider thermal conduction, compressive viscosity and optically thin radiation as damping mechanisms for MHD waves to derive a general dispersion relation and point out the flaw in the derivation of Kumar and Kumar (2006).

Recently Kumar and Kumar (2006) have investigated the damping of MHD waves in prominences, PCTRs and the corona. They have adopted the same technique as of Porter, Klimchuk , and Sturrock (1994), but their dispersion relation is a fifth-order polynomial. We, therefore, re-derive the dispersion relation taking account of thermal conduction, compressive viscosity and optically thin radiation and find it to be a sixth-order polynomial.

The relevant MHD equations in the homogeneous plasma are:

$$\frac{\partial \rho}{\partial t} + \nabla \cdot (\rho \mathbf{v}) = 0, \tag{1}$$

$$\rho \frac{D\mathbf{v}}{Dt} = -\nabla p + \frac{(\nabla \times \mathbf{B}) \times \mathbf{B}}{4\pi} - \nabla \cdot \mathbf{\Pi}, \tag{2}$$

$$\frac{D\mathbf{B}}{Dt} = \nabla \times (\mathbf{v} \times \mathbf{B}), \tag{3}$$

$$\frac{Dp}{Dt} + \gamma p (\nabla \cdot \mathbf{v}) + (\gamma - 1)\rho \, L(\rho, T) - (\gamma - 1)\nabla \cdot (\kappa \nabla T) = 0, \tag{4}$$

$$p = \frac{R}{\mu}\rho T = \frac{2\rho}{m_p} k_B T, \tag{5}$$

where $\rho$, $n$, $\mathbf{v}$, p, $\mathbf{B}$, $\gamma, R, \mu$ and T respectively are the total mass density, electron number density, velocity, total pressure, magnetic field vector, ratio of specific heats, gas constant, mean molecular weight of gas and temperature. $\Pi$ is viscosity tensor, $\kappa$ is thermal conductivity tensor, $k_B$ is Boltzmann constant, $m_p$ is proton mass, $L(\rho,T)$ is the net heat loss function per unit mass and time having the form $L(\rho,T) = \chi \rho^2 T^\alpha - h\rho^a T^b$, where $\chi$ and $\alpha$ are temperature - dependent piecewise continuous function (Hildner, 1974). h, a, and b are constant which modify the heating term (cf., Dahlburg and Mariska, 1988; Carbonell, Oliver, and Ballester, 2004). Following Porter, Klimchuk, and Sturrock (1994) we consider uniform background magnetic field, $B_0$ directed along the z-axis and homogeneous background plasma, with constant equilibrium values $\rho_0$, $T_0$, $p_0$, and $v_0 = 0$. We note that Kumar and Kumar (2006) have described the same procedure as of Porter, Klimchuk, and Sturrock (1994). When we linearize Equations (1) – (5) under the first –order approximation, assuming all disturbances in terms of Fourier components, $\exp(i\mathbf{k}\cdot\mathbf{r} - i\omega t)$, where $\mathbf{k} = k_x \hat{x} + k_z \hat{z}$, we derive the following dispersion relation :

$$\omega^6 + iA\omega^5 - B\omega^4 - iC\omega^3 + D\omega^2 + iE\omega - F = 0, \qquad (7)$$

where, $\quad A = 2c_0 + c_1$,

$\quad B = (c_s^2 + v_A^2)k^2 + c_0(2c_1 + c_0)$,

$\quad C = c_0(k^2(c_s^2 + 2v_A^2) + c_0 c_1) + c_2 k^2 + c_3$,

$\quad D = c_0\left(c_2 k^2 + v_A^2 k^2 c_0 + c_6\right) + c_2 c_4 + c_s^2 c_5$,

$\quad E = c_2 c_5 + c_0\left(c_2 c_4 + c_s^2 c_5 + c_0 c_5 c_7\right)$,

$\quad F = c_0 c_2 c_5$.

and $\quad c_0 = \dfrac{1}{p_0} A T_0$,

$\quad c_1 = \dfrac{1}{3\rho_0}\eta_0(k_x^2 + 4k_z^2)$,

$\quad c_2 = \dfrac{1}{\rho_0}(AT_0 - H\rho_0)$,

$$c_3 = \frac{1}{3\rho_0}\eta_0 k_z^2 (4v_A^2 k^2 + 9c_s^2 k_x^2),$$

$$c_4 = \frac{1}{\rho_0} 3\eta_0 k_x^2 k_z^2,$$

$$c_5 = v_A^2 k^2 k_z^2,$$

$$c_6 = \frac{1}{3\rho_0}\eta_0 k_z^2 (8v_A^2 k^2 + 9c_s^2 k_x^2),$$

$$c_7 = \frac{1}{3\rho_0} 4\eta_0.$$

where,  $A = (\gamma-1)(\kappa_\| k_z^2 + \rho_0 L_T)$,  $H = (\gamma-1)(L + \rho_0 L_\rho)$,  $c_s^2 = \frac{\gamma p_0}{\rho_0}$,

$v_A^2 = \frac{B_0^2}{4\pi\rho_0}$, $L_\rho = \left(\frac{\partial L}{\partial \rho}\right)_T$, $L_T = \left(\frac{\partial L}{\partial T}\right)_\rho$ and $\eta_0 = 10^{-17} T_0^{5/2}$ and $\kappa_\| = 10^{-11} T_0^{5/2}$

(cf., Braginskii, 1965; Porter, Klimchuk, and Sturrock, 1994). Here all units are considered in MKS.

Introducing new variables $k_T'$ and $k_\rho$ as defined by Field (1965), we have

$$k_\rho = \frac{\mu(\gamma-1)\rho_0 L_\rho}{Rc_s T_0} \quad \text{and} \quad k_T' = k_T + \frac{k_\|^2}{k_{k_\|}} + \frac{k_\perp^2}{k_{k_\perp}}; \quad k_T = \mu(\gamma-1)L_T/Rc_s,$$

$k_k = Rc_s\rho_0/\mu(\gamma-1)k$ where $k_T$ and $k_\rho$ are the wavenumbers of sound waves whose angular frequencies are numerically equal to growth rates of isothermal and isochoric perturbations respectively. The sign depends on the derivatives of $L(\rho, T)$; $k_k$ is the reciprocal of the mean free path of the conducting particles. In the presence of magnetic field, $k_T'$ corresponds to $k_T$ modified by conduction effects.

Using the expressions for A and H, $k_\rho$ and $k_T'$ become

$$k_\rho = \frac{H\rho_0}{p_0 c_s} \quad \text{and} \quad k_T' = \frac{AT_0}{p_0 c_s}.$$

In terms of these new variables, Equation (7) can be written as:

$$\omega^6 + iA_1\omega^5 - B_1\omega^4 - iC_1\omega^3 + D_1\omega^2 + iE_1\omega - F_1 = 0 \qquad (8)$$

where $A_1 = 2c_s k_T^/ + c_1,$

$$B_1 = (c_s^2 + v_A^2)k^2 + k_T^/ c_s (2c_1 + k_T^/ c_s),$$

$$C_1 = k_T^/ c_s \left[ k^2(c_s^2 + 2v_A^2) + k_T^/ c_s c_1 \right] + \frac{(k_T^/ - k_\rho)}{\gamma} c_S^3 k^2 + c_3,$$

$$D_1 = k_T^/ c_s \left( \frac{(k_T^/ - k_\rho)}{\gamma} c_S^3 k^2 + v_A^2 k_T^/ c_s k^2 + c_6 \right) + \frac{(k_T^/ - k_\rho)}{\gamma} c_S^3 c_4 + c_S^2 c_5,$$

$$E_1 = \frac{(k_T^/ - k_\rho)}{\gamma} c_S^3 c_5 + k_T^/ c_S \left( \frac{(k_T^/ - k_\rho)}{\gamma} c_S^3 c_4 + c_S^2 c_5 + c_5 c_7 c_s k_T^/ \right),$$

$$F_1 = k_T^/ c_s \frac{(k_T^/ - k_\rho)}{\gamma} c_S^3 c_5.$$

We solve our dispersion relation (8) numerically with magnetic field strength of 5 G and propagation angle $\pi/4$ for coronal regime i. e., $N_0 = 5\times 10^{13}\ m^{-3}$, $T_0 = 10^6$ K, $\chi = 1.97\times 10^{24}$ W m$^3$, and $\alpha = -1$, (Hildner, 1974) which is shown in Figure 1. We get two thermal modes rather than one as reported by Kumar and Kumar (2006), which is further discussed below.

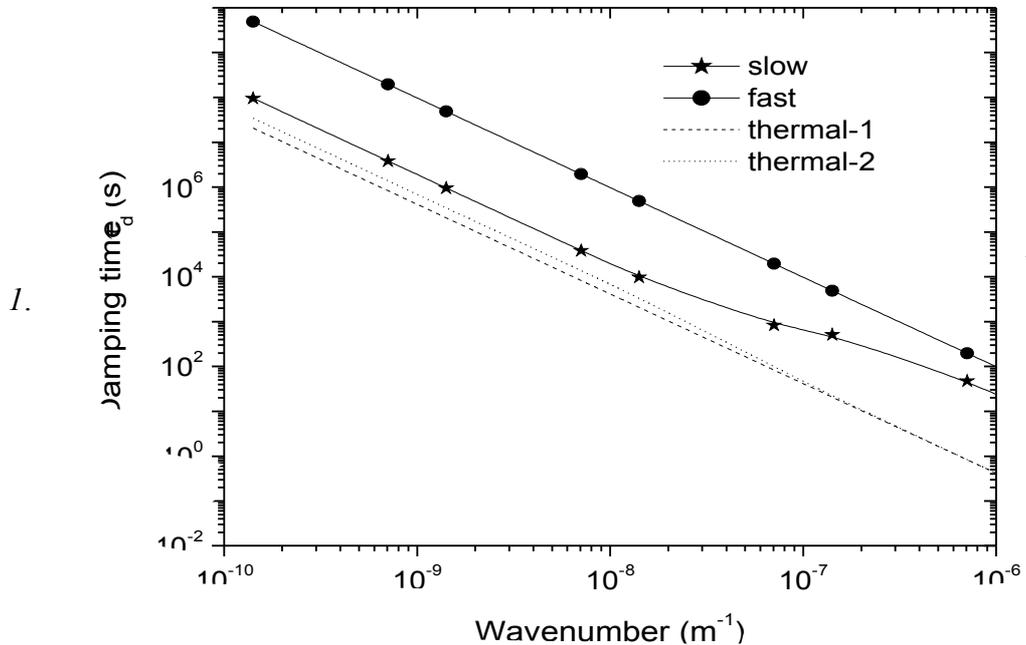

Figure 1.

Damping time as a function of wavenumber for slow, fast and thermal mode waves in coronal regime with heating mechanism characterized by a = 0 and b = 0.

We note that Kumar and Kumar (2006) have derived the dispersion relation by taking 3×3 determinant of the coefficients equal to zero. These coefficients are associated with three independent variables namely, $v_{1x}$, $v_{1z}$, and $p_1$. This will result in a dispersion relation of fifth-order polynomial. We will then have five roots out of which two correspond to slow mode and the other two correspond to fast mode. The remaining root being imaginary in nature corresponds to thermal mode. The disadvantage of this approach is that one cannot get the inequality conditions $v_{1z} \gg v_{1x}$ (for slow mode waves) and $v_{1x} \gg v_{1z}$ (for fast mode waves) on which weak damping approximation is valid (see Porter, Klimchuk, and Sturrock, 1994). It can be achieved only when we have two sets of equations in terms of two independent variables $v_{1x}$ and $v_{1z}$. In this way, a complete wave equation can be expressed in terms of a single velocity vector **v.** If we substitute the value of $p_1$ from linearized equation of energy in linearized equation of momentum using equation of continuity and equation of state, then we get two sets of equations in two independent variables $v_{1x}$ and $v_{1z}$. This reduces the number of independent variables from three to two. That means, we have 2×2 determinant of the coefficients equal to zero, which results in a dispersion relation of sixth-order polynomial instead of fifth-order polynomial as reported by Kumar and Kumar (2006). The sixth-order dispersion relation has the dissipative terms of viscosity, thermal conductivity and optically thin radiation. Solution of this dispersion relation provides six roots namely, $\omega_{1r} - i\omega_{1i}$, $-\omega_{1r} - i\omega_{1i}$, $\omega_{2r} - i\omega_{2i}$, $-\omega_{2r} - i\omega_{2i}$, $\omega_{3r} - i\omega_{3i}$ and $-\omega_{4r} - i\omega_{4i}$, where $\omega_{3r}$ and $\omega_{4r}$ are negligibly small compared to $\omega_{1r}$ and $\omega_{2r}$. Thus two roots are purely imaginary which correspond to thermal mode and the other four roots are in pair form. One pair corresponds to slow mode and the other pair to fast mode. If we consider the thermal conductivity only, we get six roots i.e., $\omega_{1r} - i\omega_{1i}, -\omega_{1r} - i\omega_{1i}, \omega_{2r} - i\omega_{2i}, -\omega_{2r} - i\omega_{2i}, \omega_{3r} - i\omega_{3i}$ and $-\omega_{4r} - i\omega_{4i}$. This means, we have slow mode, fast mode and thermal mode. When we consider the viscosity term only, we get four roots i.e., $\omega_{1r} - i\omega_{1i}, -\omega_{1r} - i\omega_{1i}, \omega_{2r} - i\omega_{2i}$ and $-\omega_{2r} - i\omega_{2i}$. This simply means that the

46

thermal mode is excited only when thermal conductivity is present in the dispersion relation.

In conclusion, we find that a general dispersion relation is a sixth-order polynomial. The flaw in the fifth-order dispersion relation by Kumar and Kumar (2006) is because of their inconsistent approach of considering three independent variables rather than two in their derivation.

## Acknowledgements

V.S. Pandey acknowledges the financial support from the CSIR, New Delhi.## References

Braginskii, S.I. : 1965, *Rev. Plasma Phys.* **1**, 205.

Carbonell, M., Oliver, R., and Ballester, J. L. : 2004, *Astron. Astrophys.* **415**, 739.

Dahlburg, R. B. and Mariska, J. T. :1988, *Solar Phys.* **117**, 511.

De Moortel, I. and Hood, A. W. : 2003, *Astron. Astrophys.* **408**, 755.

Field, G. B. :1965, *Astrophys. J.* **142**, 531.

Hildner, E. : 1974, *Solar Phys.* **35**, 123.

Kumar, N. and Kumar, P. : 2006, *Solar Phys.* 236, 137.

Porter, L. J., Klimchuk, J. A., and Sturrock, P. A. : 1994, *Astrophys. J.* **435**, 482.